# Plasmon Enhanced Quantum Properties of Single Photon Emitters with Hybrid Hexagonal Boron Nitride Silver Nanocube Systems

Mohammadjavad Dowran,[1] Andrew Butler,[2] Suvechhya Lamichhane,[3] Adam Erickson,[1] Ufuk Kilic,[2] Sy-Hwang Liou,[3] Christos Argyropoulos,[2,4,&] and Abdelghani Laraoui[1,3,*]

[1]Department of Mechanical & Materials Engineering, University of Nebraska-Lincoln, 900 N 16th St. W342 NH. Lincoln, NE 68588, USA

[2]Department of Electrical and Computer Engineering, University of Nebraska-Lincoln, 844 N. 16th St, Lincoln, NE, 68588-1212, USA

[3]Department of Physics and Astronomy and the Nebraska Center for Materials and Nanoscience, University of Nebraska-Lincoln, 855 N 16th St, Lincoln, Nebraska 68588, USA

[4]Department of Electrical Engineering, The Pennsylvania State University, 203 Electrical Engineering East, University Park, PA, 16802, USA

[&] cfa5361@psu.edu , [*]alaraoui2@unl.edu

**Abstract**

Hexagonal boron nitride (hBN) has emerged as a promising ultrathin host of single photon emitters (SPEs) with favorable quantum properties at room temperature, making it a highly desirable element for integrated quantum photonic networks. One major challenge of using these SPEs in such applications is their low quantum efficiency. Recent studies have reported an improvement in quantum efficiency by up to two orders of magnitude when integrating an ensemble of emitters, such as boron vacancy defects in multilayered hBN flakes embedded within metallic nanocavities. However, these experiments have not been extended to SPEs and are mainly focused on multiphoton effects. Here, we study the quantum single photon properties of hybrid nanophotonic structures composed of SPEs created in ultrathin hBN flakes coupled with plasmonic silver nanocubes. We demonstrate > 200% plasmonic enhancement of the SPE properties, manifested by a strong increase in the SPE fluorescence. Such enhancement is explained by rigorous numerical simulations where the hBN flake is in direct contact with the Ag nanocubes that cause the plasmonic effects. The presented strong and fast single photon emission obtained at room-temperature with a compact hybrid nanophotonic platform can be very useful to various emerging applications in quantum optical communications and computing.

## 1. Introduction

Single-photon emitters (SPEs) in solid-state platforms have attracted a large interest over the last decades for various applications in quantum information processing and quantum sensing.[1] Significant developments led to the discovery of a variety of atom-like SPEs in solid-state materials, such as semiconductor quantum dots (*e.g.*, Gallium Arsenide (GaAs) grown by molecular beam epitaxy)[2] and defect-related color centers in wide bandgap materials, *e.g.*, nitrogen vacancy (NV) centers in diamond,[3–8] and divacancies (4H, 6H) in silicon carbide.[9–12] Although substantial recent progress led to understanding and utilizing the quantum properties of SPEs, further advances are severely limited by difficulties in achieving the exact placement of quantum emitters, a weak light collection due to the high refractive index of bulk substrates, slow emission dynamics, and large-scale integration into photonic nanostructures.[4] Several candidates have been studied as sources for SPEs including tungsten diselenide ($WSe_2$),[13] molybdenum disulfide ($MoS_2$),[14] quantum dots,[2] and color centers in wide bandgap semiconductors.[15] Among these sources, hexagonal boron nitride (hBN) has emerged as a promising host of SPEs[16–23] that display promising quantum properties such as narrow emission linewidth at ambient conditions[17] and optically detected magnetic resonance (ODMR),[24] making it a promising material for integrated quantum photonics[1,25,26] and sensing.[27] SPEs have been generated in a large selection of hBN hosts including exfoliated flakes from bulk substrates,[17,18] strain-originated defects in thin hBN flakes,[28] chemical vapor deposition (CVD) grown crystals,[29–31] and commercially available nanoflakes.[32] Various processing techniques have been used to increase the density of SPEs in hBN and improve their quantum properties including annealing at high temperatures (> 850 $^0$C) under Ar and/or $O_2$ flow,[33] nanoindentation with atomic force microscopy (AFM),[34] plasma treatment,[20] and ion beam implantation of different species (C, O, N).[35] While in most cases, the density of observed SPEs is high (≥ 0.5 emitter in 1 μm$^2$ area) with narrow emission spectrum, their quantum efficiency is still low with a fluorescence emission rate way below 1 x 10$^6$ c/s.[33]

Different approaches have been used to increase the fluorescence rates of SPEs in diamond nanocrystals and films by coupling them deterministically with photonics crystals,[36,37] bullseye grating,[38,39] and plasmonic nanocavities.[40] However, these methods remain challenging to be applied to hBN due to the difficulty in creating SPEs in hBN nanoflakes[41] with desired spectral properties that match the cavity optical frequency modes. Kim et al. created SPEs on top of hBN photonic crystal by annealing at 850 °C, coupled them spatially within the cavity, and did not see any enhancement in their quantum properties due to the off spectral overlap between the SPEs and the cavity frequency modes.[42] A different approach was used very recently by spreading silver nanocubes (SNCs) on top of hBN flakes containing high density boron vacancy ($V_B^-$) emitters deposited on thin metal (Au and Ag) films.[43,44] An overall intensity enhancement of up to 250 times was obtained with a high signal to noise ratio (SNR) ODMR signal[43,44] mainly due to the large Purcell enhancement of the plasmonic Ag nanoantennas.[45–47] While these findings are very promising to quantum sensing,[48,49] these experiments were mainly focused on multiphoton effects, such as fluorescence enhancement, and the single photon enhancement from coupling SPEs in hBN flakes with SNCs has not been reported yet.

In this work, we couple SPEs in ultrathin hBN flakes (thickness of 15 nm) to the dominant plasmonic mode of SNCs (size of 98 nm) at room temperature. We experimentally observe a narrowing of the SPEs emission spectra combined with a reduction in their fluorescence lifetime that is accompanied by an enhancement of their single photon fluorescence rate at ambient conditions. Our results provide a first step towards achieving a major speedup and enhancement in the single photon emission of compact nanophotonic devices that can be used in emerging quantum optical applications.[50]

## 2. Methodology
### 2.1. Sample Preparation and Creation of SPEs in hBN flakes
hBN flakes are exfoliated from bulk hBN crystals (hq graphene) using vinyl tape[33] and transferred to a marked 200 nm thick $SiO_2$/Si substrate. To create a high density (> 0.5 SPE/1μm$^2$) of stable SPEs we followed the recipe described in reference [33] by annealing the hBN flakes on $SiO_2$/Si substrate under $O_2$ flow (950 sccm) at a temperature of 1100 $^0$C for 4 hours. It is believed that oxygen molecules may break down and react with nitrogen or boron atoms to form different defect species (e.g., oxygen impurities, substitutional oxygen at N or B sites) or take part in the etching process to form optically active vacancy related defects.[33,51,52] The hBN flakes were scanned prior to the annealing process to ensure the origin of the defects is not related to the strain variation across the flakes.

We imaged optically different grids on the SiO$_2$/Si substrate after the annealing process (Supporting Information Section I) to identify and select only ultrathin hBN flakes (< 20 nm in thickness). We measured their thickness by using an AFM setup (Supporting Information Section III). To enhance the quantum properties of SPEs on a selected thin hBN flake we spin coated commercial SNCs with a size of 98 nm ± 7 nm (nanoCopisix) on the annealed SiO$_2$/Si substrate (**Fig. 1a**). We optimized the spin coating parameters to increase the chance of getting the SNCs coupled spatially with the SPEs (Supporting Information Section V). We used a home-built confocal fluorescence microscope (**Fig. 1b**) at room temperature to characterize the quantum properties of SPEs in the hBN flakes before and after depositing the SNCs. The setup has four modalities: optical/fluorescence imaging to isolate thin flakes with SPEs, spectroscopy to check the SPE emission wavelength, anti-bunching $g^{(2)}$ experiments to check whether the emitters in hBN are indeed SPEs, and lifetime measurements to measure the SPEs fluorescence lifetime.

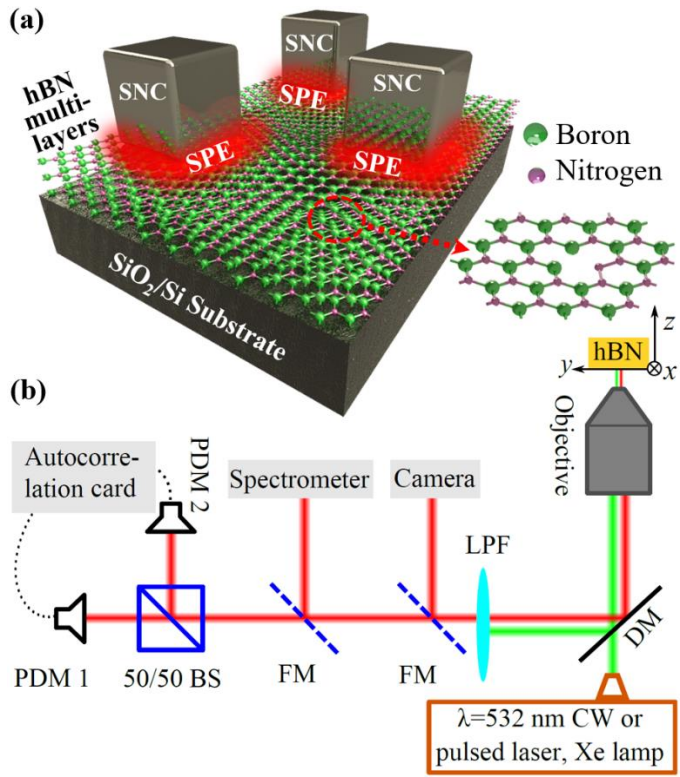

**Figure 1.** **(a)** Enhanced quantum properties based on hybrid nanophotonic structures composed of SPEs in hBN combined with localized plasmon excitations from SNCs. Insert of **(a)**: a sketch of the distribution of hBN flake atoms, composed of boron (green) and nitrogen (purple) atoms in a 2D lattice. **(b)** A schematic of our confocal fluorescence microscope setup consisting of a CCD camera to isolate the hBN flakes, a spectrometer, auto-bunching $g^{(2)}$ setup, lifetime measurements, and fluorescence imaging. DM: dichroic mirror, CW: continuous wave, LPF: long-pass filter, FM: flipped mount mirror, BS: beam splitter.

### 2.2. Optical characterization

To locate the hBN flakes and identify their thickness based on their color[53] a white light (Xe lamp) is focused on the SiO$_2$/Si substrate and the reflected light is then focused on a CCD camera, as depicted in **Fig. 1b**. We used two types of 532-nm lasers: A CW laser (MLL-FN-532-1000 mW) to perform fluorescence imaging and $g^{(2)}$ measurements, and a 10 ps pulsed laser (SuperK Fianium FIU-15 from NKT Photonics) to measure the fluorescence lifetime. The 532 nm laser is reflected from a dichroic mirror (Semrock, FF560-FDi01) and focused on the hBN flakes using a Nikon oil objective (NA= 1.25). The fluorescence (560-800 nm) emitted from the hBN flake is collected by the same objective, transmitted through the dichroic mirror, and sent to the detection setup to characterize the SPEs quantum properties using flip-mount mirrors. The fluorescence is then focused to a balanced split-single mode fiber (Thorlabs Model TW670R5F1) using 0.25 NA objective and coupled to two single photon detection modules (PDM, Micro Photon Devices). A notch filter (Semrock, NF01-532U-25) is placed in the fluorescence path to block completely the 532-nm excitation light.

The fluorescence imaging of the hBN flake is performed by scanning the $SiO_2$/Si sample (travel range of 100 μm along x, y, and z) mounted on a closed loop three axis nanopositioning linear stage (NPXYZ100SG, Newport) and collecting the emitted photons from a confocal spot (~ 320 nm) by using one of the PDM modules. **Fig. 2a** shows the fluorescence image from the lower corner (10 μm x 10 μm) of the hBN flake in the lower insert of **Fig. 2a** with a high density of emitters (~ 0.5 μm$^2$). The thickness of the flake is ~15 nm, determined from AFM measurements (**Fig. S3**). The fluorescence spectrum from each emitter in **Fig. 2a** is recorded by using Solis Triax 320 spectrometer coupled to Andor camera (iDus 420 CCD). The fluorescence is focused by 75 mm achromatic lens (Thorlabs AC254-075-A) on the spectrometer focal plane and collected through a slit of 50 μm. **Fig. 2b** displays the normalized fluorescence spectra (subtracted from the substrate background) obtained from two selected emitters SPE 9 and SPE 10 in the hBN flake. Sharp emission lines with full width at half-maximum (FWHM) of 2.94 nm and 6.3 nm are observed by the emitters at 686.5 nm and 616 nm, respectively. We attribute these peaks to the zero-phonon-line (ZPL) of the defects.[18] In **Fig. 2c** we plot the histogram of the ZPL peaks occurrence as function of wavelength for 16 SPEs (see **Table S1**) from the hBN flake in **Fig. 2a** and found that all emitters are centered in a spectral window of 600-710 nm with a negligible phonon side band (PSB). This correlates well with recent studies done on SPEs created by annealing hBN flakes under $O_2$ flow.[33]

To check the single-photon emission properties of the emitters in **Fig. 2a**, we performed anti-bunching g$^{(2)}$ measurements in the Hanbury Brown and Twiss (HBT) configuration. We used a time-tagger device (PicoHarp 300) to measure the autocorrelation g$^{(2)}$ response from the photons collected by the two PDM modules as a function of the time delay *t* between the two detectors (±20 ns). We found > 90% of the emitters in **Fig. 2a** have a dip of g$^{(2)}$< 0.25 (**Fig. 2d**, and **Table S1**), a clear indication of SPEs.[47] For short time scales (t ≲ 20 ns), the g$^{(2)}$ response of the SPEs (e.g., emitter 9 in **Fig. 2d**) is normalized and fitted to the following equation: $g^{(2)}(t) \simeq 1 - (1 + a_1)e^{-t/\tau_1}$, where *a$_1$* is the anti-bunching factor and $\tau_1$ is the excited state lifetime which includes both the radiative and nonradiative transition lifetimes.[54,55] We obtain *a* = -0.47 ± 0.05 and $\tau_1$ = 2.9 ns ± 0.15 ns by fitting the measured curve in **Fig. 2d** (solid line).

To measure the excited-state lifetime of the emitters, we used a 10 ps pulsed green laser to excite the emitters and time-tagger device (PicoHarp 300) to perform time dependent photon statistics. For this experiment, the time tagger measures the time difference between the excitation 10 ps pulse and the detected SPE photons of the PDM detector. We calibrated the lifetime setup and found a response function (IRF) of 19 ps. The lifetime signal is normalized and found to be fitted with only one exponential model using the following equation: $a_2 e^{-t/\tau_3}$, where *t* is the varying time, *τ$_3$* is the fluorescence lifetime, and *a$_3$* a weighing factor.[18] The lifetime *τ$_3$* of SPE 9 is ~ 4.5 ns ± 0.23 ns (**Fig. 2e**) and varies from 1 ns to 5 ns for the 16 SPEs studied in the hBN flake in **Fig. 2a** (see **Table S1**).

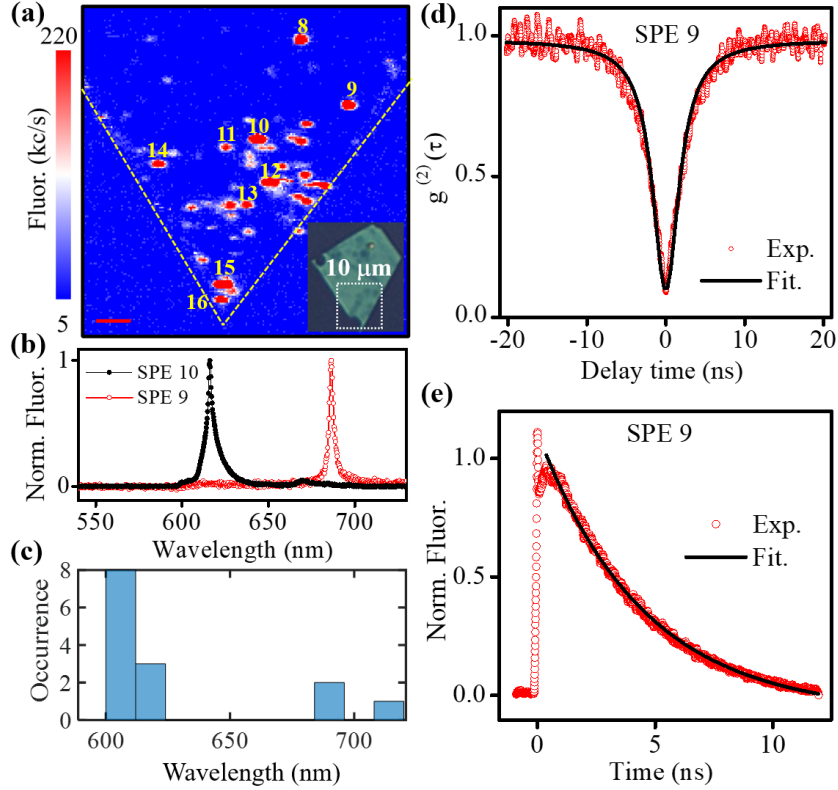

**Figure 2.** *Characterization of the SPEs quantum properties along a selected 15 nm thick hBN flake.* **(a)** 10 μm x 10 μm fluorescence image of the hBN flake lower part shown in the inset (optical image). Red spots indicate high photon count rates emitted from the flake, where 16 spots were labeled and characterized. **(b)** Spectrum of the fluorescence intensity vs wavelength of two selected emitters SPE 9 and SPE 10 marked in **(a)**. **(c)** Measured emission wavelength distribution of the 16 SPEs characterized in the hBN in **(a)**. **(d)** Autocorrelation $g^{(2)}$ response of emitter SPE 9 as function of the delay time between the two PDM modules. **(e)** Lifetime response of emitter SPE 9 as function of the time between the 532-nm laser excitation 10 ps pulse and the detected fluorescence in one of the PDM modules. The scale bar in **(a)** is 1 μm.

## 3. Results and Discussion
### 3.1. Plasmon enhancement of the SPEs quantum properties in a hybrid hBN-SNC system

We measured the SPE optical properties before and after spin-coating of 98 nm ± 7 nm SNCs on top of the hBN flake depicted in the insert of **Fig. 2a**. By using an overlay method and by verifying the spectra of the emitting spots, we identified the location of the labeled SPEs after spin coating of the SNCs. We detail the distribution of the spin coated SNCs on top of the hBN flake in Supporting Information Section V. We verified the distribution of the SNCs on the hBN flake by AFM analysis of the SNCs (**Fig. 3a**) and confirmed the size of the cubes of 98 nm ± 7 nm (**Fig. 3b**). Moreover, the plasmonic spectral mode of the silver nanocube supports only a selection of emission modes from the SPEs in the hBN flake in the spectral region around the nanocube's resonance frequency. Hence, not all emission modes from SPEs get enhanced and, therefore, the wavelength response of SPEs gets narrower, as was demonstrated before in the photoluminescence spectrum of a relevant nanocube cavity system combined with $MoS_2$.[56] This effect is also present in our system and depicted in **Fig. 3c** with a decrease in the FWHM of the ZPL peak by 30%. We demonstrate a plasmonic enhancement of the quantum properties of emitter SPE 16 manifested by a narrowing of the g$^{(2)}$ response (**Fig. 3d**), and a decrease in the fluorescence lifetime from 3.13 ns ± 0.16 ns to 1.56 ns ± 0.08 ns (**Fig. 3e**). The plasmon SPE enhancement rate $\frac{\frac{1}{\tau_{w.SNC}}}{\frac{1}{\tau_{w/o\ SNC}}} = \frac{\tau_{w/o\ SNC}}{\tau_{w.\ SNC}} = 2$ corresponding to an enhancement rate of 200%. We attribute this to the Purcell effect that leads to a radiative rate enhancement and the lossy nature of the plasmonic silver nanocube structure which increases the non-radiative rate. Both effects are related to the local confinement of the electric field at the SNC surface,[45,46,57] confirmed by COMSOL modeling (see discussion below). The excited state lifetime reduction is dominated by the increase in the efficiency of the defect radiative decay, confirmed by measuring the measured decay rate enhancement with the saturated intensity enhancement (below).[58] The enhancement factor varies across SPEs. For example, in SPE 9 there is no enhancement and in SPE 15 we observe an enhancement factor of 70% (**Fig. S6**). We explain this variation by the non-spatial (e.g., SPE 10) and/or non-spectral (e.g., SPE 9) overlap between the SPE and the SNC localized plasmon modes in the range of 450-650 nm computed and presented in **Fig. 4a**.

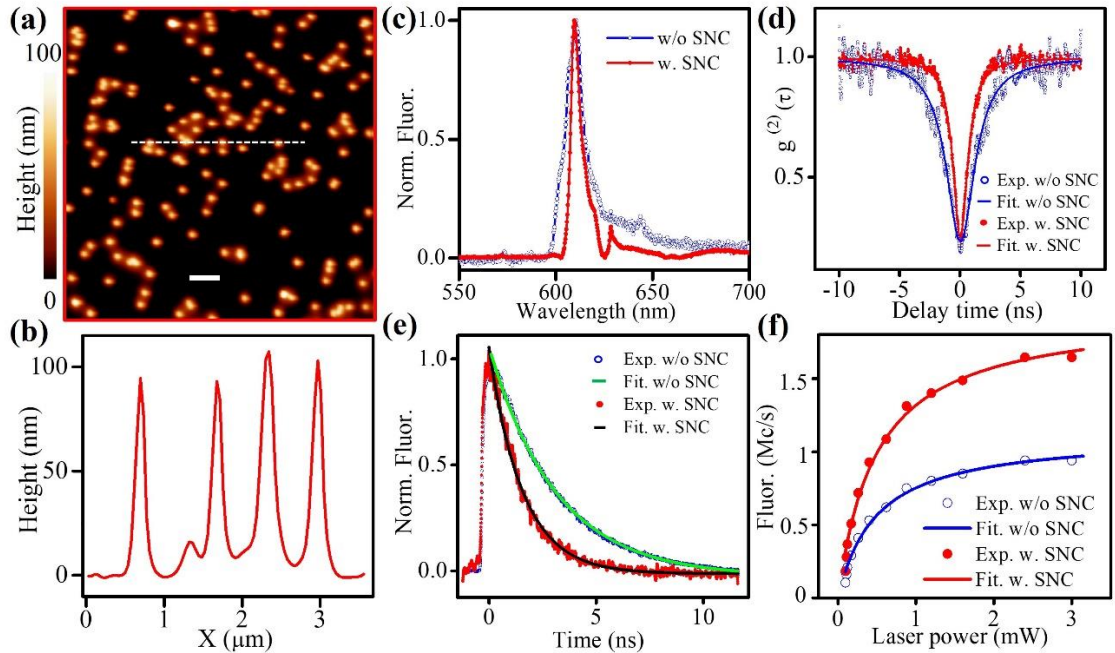

**Figure 3.** *Plasmon-assisted enhancement of SPEs quantum properties*. **(a)** A topography map of the SNCs spread on top of the hBN flake in the insert of **Fig. 2a**. **(b)** A transverse cross-cut (scattered line) of the AFM image in **(a)**. The scale bar in **(a)** is 0.5 μm. The SNCs have a size varying from 95 to 105 nm above the hBN flake (thickness is 15 nm). **(c)** Measured spectra **(c)**, **(d)** g$^{(2)}$ responses, and **(e)** lifetime rates of SPE 16 before (open-circle scattered) and after (filled-circle scattered) depositing the 98 nm nm SNC. **(f)** Measured and calculated fluorescence intensity of SPE 16 as function of the CW green laser power before (open-circle scattered) and after (filled-circle scattered) depositing the 98 nm SNC. The red and blue lines in **(f)** are fitted curve according to Ref. 58.

To know the saturated count rate of the emitter before and after spin coating the SNCs (size of 98 nm) we measured the fluorescence intensity of SPE 16 as function of the CW laser power and found a saturation count rate of ~ 1 x 10$^6$ c/s without SNC (open circle line in **Fig. 3f**) and a saturation count rate of ~ 2 x 10$^6$ c/s with SNC (filled-circle line in **Fig. 3f**). The saturation curves for the SPE 16 with and without SNC are fitted (**Fig. 3f**) to:[58] $I = \frac{I_\infty P}{(P_{sat}+P)}$, where $I_\infty$ is the saturated count rate and $P_{Sat}$ is the saturation power. $I_{\infty,w/o\,SNC} = 0.97 \times 10^6$ c/s and $I_{\infty,w.SNC} = 1.7 \times 10^6$ c/s for the uncoupled and coupled SPE respectively. The enhancement factor $\frac{I_{\infty,SNC}}{I_{\infty,w/o\,SNC}}$ is 1.75, which is dominated by the enhancement of the radiative decay channel.[58] The saturation power curve of the coupled SPE-SNC system is subtracted from the fluorescence emission power dependence of an isolated SNC away from the emitter on the hBN flakes (**Fig. S5**). Therefore, the saturation curve for the SPE 16 with SNC is assumed to originate solely from the plasmonic-enhanced fluorescence of the SPE. The difficulty in measuring the SPE saturation count rate after depositing the nanocubes arises from the high intensity autofluorescence generated by the 98 nm SNCs in the 600-700 nm spectral window (up to $5 \times 10^6$ c/s at saturation, **Fig. S5**). However, an increase in the SNR of the SPE 16 with SNC fluorescence is clearly depicted in **Fig. 3c** and **Fig. 3d** where the measurements are taken at the same averaging time before and after spin coating the SNCs.

### 3.2. Modeling the plasmonic modes of the SNC-hBN system

The finite-element method simulation software (COMSOL Multiphysics®) was used to accurately model the scattering response of the SNCs. The SNC was encased in a spherical domain with scattering boundary conditions to mimic an open boundary. The edges of the nanocube were 98 nm long and smoothed with a radius of 10 nm curvature to be closer to the shape of the experimental samples. All the appropriate substrate materials below the nanocube were included in our calculations. The hBN, Ag, $SiO_2$, and Si materials that compose the currently studied system were all modelled using their frequency dependent dielectric constants.[59–62] To calculate the scattering cross section, the scattered-field formulation was used which considers the analytical solution for an incident plane wave in the absence of the nanocube as the background electric field. The incident plane wave is transverse-magnetic (TM) polarized and impinges upon the cube under normal incidence. The scattering energy is calculated by measuring the power of the field scattered by the cube. This energy is then divided by the geometrical cross section of the SNC to compute the scattering cross section. Since the nanocube structure is symmetric, identical scattering results are obtained when the incident plane wave is transverse-electric (TE) polarized or incoherently polarized containing both TE and TM polarizations. **Fig. 4a** shows part of the scattering cross section as a function of the wavelength. A broad scattering plasmon resonance is observed in the range 450 nm-650 nm at the central wavelength of 550 nm. We plot the electric field enhancement factor at a wavelength of 610 nm from the 98 nm SNC in **Fig. 4b** (side view) and Fig. **4c** (top view) and demonstrate an increase of up 28 times at the edges of the cube. Since the SPE location across the hBN thickness cannot be determined experimentally by optical measurements, we integrated the overall field enhancement as a function of the distance *d* from the silver cube across the hBN multilayer. We observe an overall field enhancement of 3 (300%) at the interface of hBN and SNC reduced to 1.6 (60%) along the interface of hBN and $SiO_2$ with results demonstrated in **Fig. 4d**. The extent of the field enhancement from a SNC is related to the evanescent waves of the plasmonic excitations localized on the surface of the SNC, and decays as the measurement plane gets farther away from the SNC, as depicted in **Fig. 4d**. This correlates well with the enhancement factor of the fluorescence lifetime and intensity measured on SPE 16 (**Fig. 3**) and SPE 15 (**Fig. S6**).

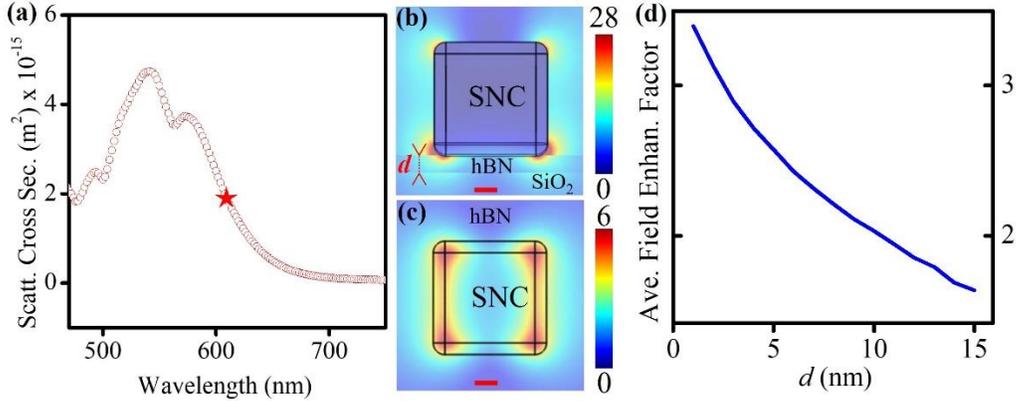

**Figure 4.** *Simulation results of the hybrid hBN-SNC system plasmonic modes*. **(a)** Computed scattering cross section spectrum that shows a higher-order resonance near 550 nm. **(b)** Side and **(c)** top views of the field enhancement factor along a 98 nm SNC demonstrating the electric field distribution of the plasmonic resonance mode at wavelength 610 nm corresponding to the wavelength of SPE 16 studied in **Fig. 3**. **(d)** The average electric field enhancement factor as a function of the distance $d$ in hBN, as sketched in **(b)**. The scale bar in **(b)** and **(c)** is 20 nm.

## 4. Conclusions

In summary, we demonstrated the coupling of silver nanocube plasmonic modes (98 nm in size) with SPEs in layered 15 nm thick hBN flakes. An overall enhancement of the SPE fluorescence lifetime and intensity of up to 200% is obtained, corresponding to a saturation count rate $> 2 \times 10^6$ c/s. We confirmed the measured plasmonic enhancement of the SPE properties by using finite-element method simulations to accurately model the scattering response of the hybrid SPE-SNE system. The presented strong and fast single photon emission obtained at room-temperature with a hybrid nanophotonic platform can be very useful to various emerging applications in quantum optical communications and computing. To better improve the spatial and spectral overlap of the SPEs in hBN with the optical frequencies of the metal nanocavities a combination of processing and nanofabrication methods must be followed. This includes a deterministic creation of SPEs in hBN flakes deposited on epitaxial metal films (high reflectivity) by for example nanoindentation with AFM,[34] followed by precise nanofabrication of dielectric or metal nanostructures such as photonics crystals,[36] bullseye grating,[38,39] and plasmonic nanocavities.[40] However, even the current results prove that room-temperature solid-state quantum emitters in hBN or other two-dimensional (2D) van der Waals (vdW) materials[63] can be an ideal platform for integrated quantum photonics.[1,25,26]

**Author Contributions**
M.D. synthesized the hBN flakes, performed the optical measurements and analyzed the data; A.B. performed COMSOL modeling; S.L and S.-H.L prepared the $SiO_2$/Si substrates and made the marks; A.E. assisted M.D. in the AFM measurements; U.K. helped in the designs in **Fig. 1** and performed ellipsometry measurements (not reported here) to check the optical properties of $SiO_2$/Si substrates. A.L. and C.A. designed the experiments and supervised the project. A.L. wrote the manuscript with contributions of all authors. All authors have given approval to the final version of the manuscript.

**Acknowledgements**
This material is based upon work supported by the NSF/EPSCoR RII Track-1: Emergent Quantum Materials and Technologies (EQUATE), Award OIA-2044049. The research was performed in part in the Nebraska Nanoscale Facility: National Nanotechnology Coordinated Infrastructure and the Nebraska Center for Materials and Nanoscience (and/or NERCF), which are supported by NSF under Award ECCS: 2025298, and the Nebraska Research Initiative. C.A. acknowledges partial support from the Office of Naval Research Young Investigator Program (ONR-YIP) (Grant No. N00014-19-1-2384) and the NASA Nebraska Space Grant Fellowship. We thank K. Ambal and S. Halder for helping to set up the confocal fluorescence microscope setup.

**Conflict of Interest**
The authors declare no competing financial interest.

**Data Availability Statement**
The data that support the findings of this study are available from the corresponding author upon reasonable request.

**Keywords**
Single photon emitters, hexagonal boron nitride, Plasmon, Purcell effect.

**Supporting Information**

**I. Exfoliating hBN flakes**
Silicon substrates with 200 nm thermal silicon oxide ($SiO_2$) were marked with reference grid marks by using laser lithography (DWL-66). The marked silicon substrates were then sonicated in acetone to clean them from mask residue, followed by plasma processing at 80 W under 15 sccm of oxygen flow for 10 min. The substrates have maximum acceptance of the flakes within 10 minutes after plasma processing. For exfoliating hBN flakes on the silicon substrate, we place a piece of hBN crystal (hq graphene) on a piece of vinyl tape with less residue. By folding the tape on itself multiple times we form an even distribution of hBN layer over the tape piece. The flakes can be transferred to the designated marked silicon substrate by placing the tape on the substrate and rubbing it with a soft plastic or a Q-tip. With an optical microscope, we check the distribution and thickness of the flakes on the substrate. hBN flakes of different thicknesses show distinct colors depending on their thicknesses (**Fig. S1**), which were previously characterized using an AFM analysis. To clear tape residues and increase the bonding between hBN flakes and the substrate, we place the substrate on the hot plate at 350 $^0$C for 30 min.

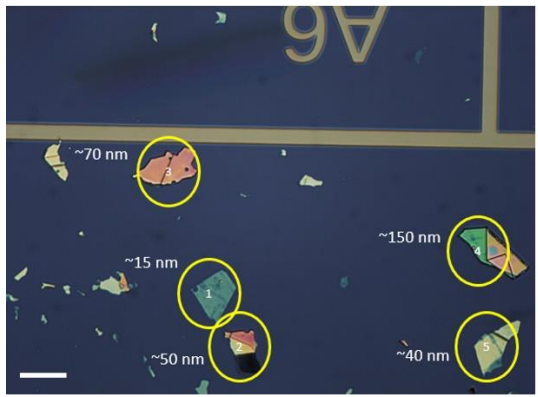

**Figure S1: (a)** Annealed hBN flakes with various thicknesses show different colors under an optical microscope. The scale bar is 30 μm. We identify the thickness of each flake with a certain color coding using combined optical and AFM measurements.

## II. Annealing of hBN flakes for SPE creation

To anneal hBN flakes on the SiO$_2$/Si substrate, we place it at the center of the annealing furnace tube with three temperature sensors (Lindberg Blue), set to provide 1100 $^0$C at the spot where the sample is placed. After turning on the vacuum pump and setting the oxygen flow to 1000 sccm, we turn on the temperature controller to increase the temperature to 120 $^0$C/min and then anneal for four hours. After this period, we let it cool down on its own and under oxygen flow. The timeline of the temperature controller settings for the annealing process is shown in **Fig. S2**. After annealing, the sample is left inside the furnace to cool down to room temperature which typically takes 4-5 hours. The graph in **Fig. S2** shows the set point on the temperature controller which we set at max 1150 $^0$C that corresponds to 1100 $^0$C inside furnace tube.

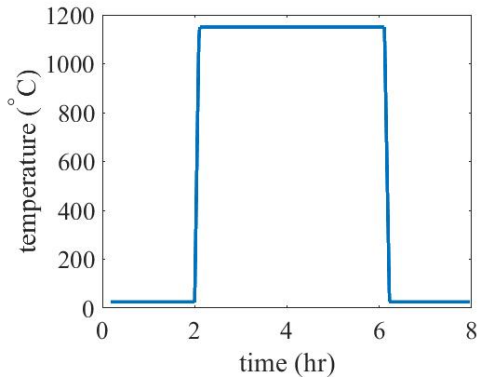

**Figure S2:** The timeline for the annealing temperature settings. The temperature increases from room temperature to 1100 $^0$C at a rate of 120 $^0$C /min.

## III. Topography measurements of the hBN flake

We measured the hBN flake (insert of **Fig. 2a** in the main paper) thickness by using atomic force microscope (AFM) Innova from Bruker and related it to their color under the optical microscope. We used AFM tips Tap300-G (Budget sensors) in the tapping mode with a force constant of 40 N/m and resonance frequency of ~ 300 kHz. In **Fig. S3a** we plot the topography image of the flake and in **Fig.S3b** we display the lateral cross lines and find a thickness of 15 nm ±2 nm.

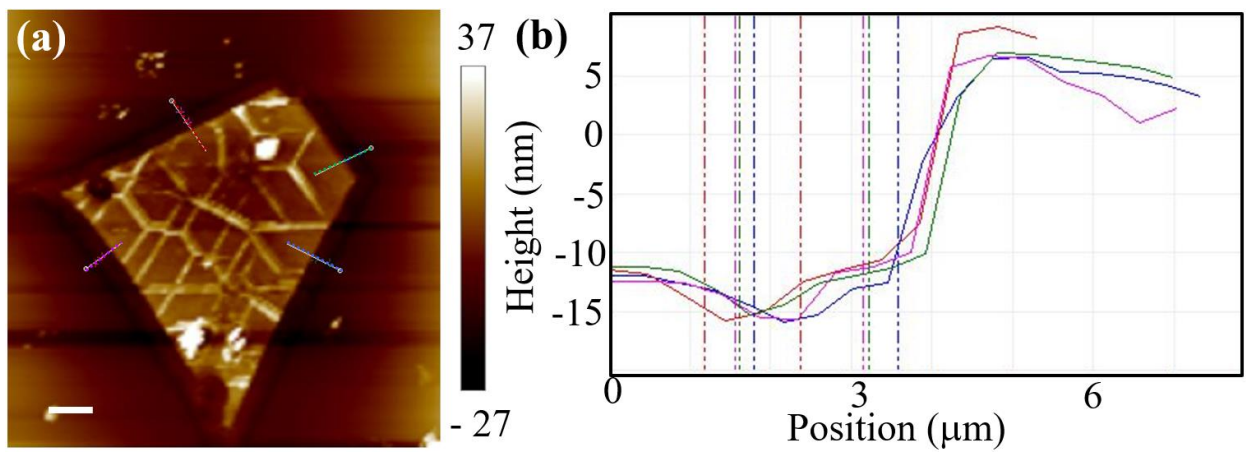

**Figure S3:** **(a)** AFM image from the hBN flake under study in the manuscript. The scale bar is 10 μm. **(b)** The height of the measurement scattered lines from the AFM image in **(a)** shows the thickness of the flake is uniformly ~15 nm ±2 nm.

## VI. Photophysical properties (spectrum, $g^{(2)}$, lifetime) of 16 emitters in flake 1

In **Table S1** we list the optical properties of 16 identified SPEs on the 15 nm thick hBN flake studied in this paper. PL is the photoluminescence and λ is wavelength.

**Table S1**

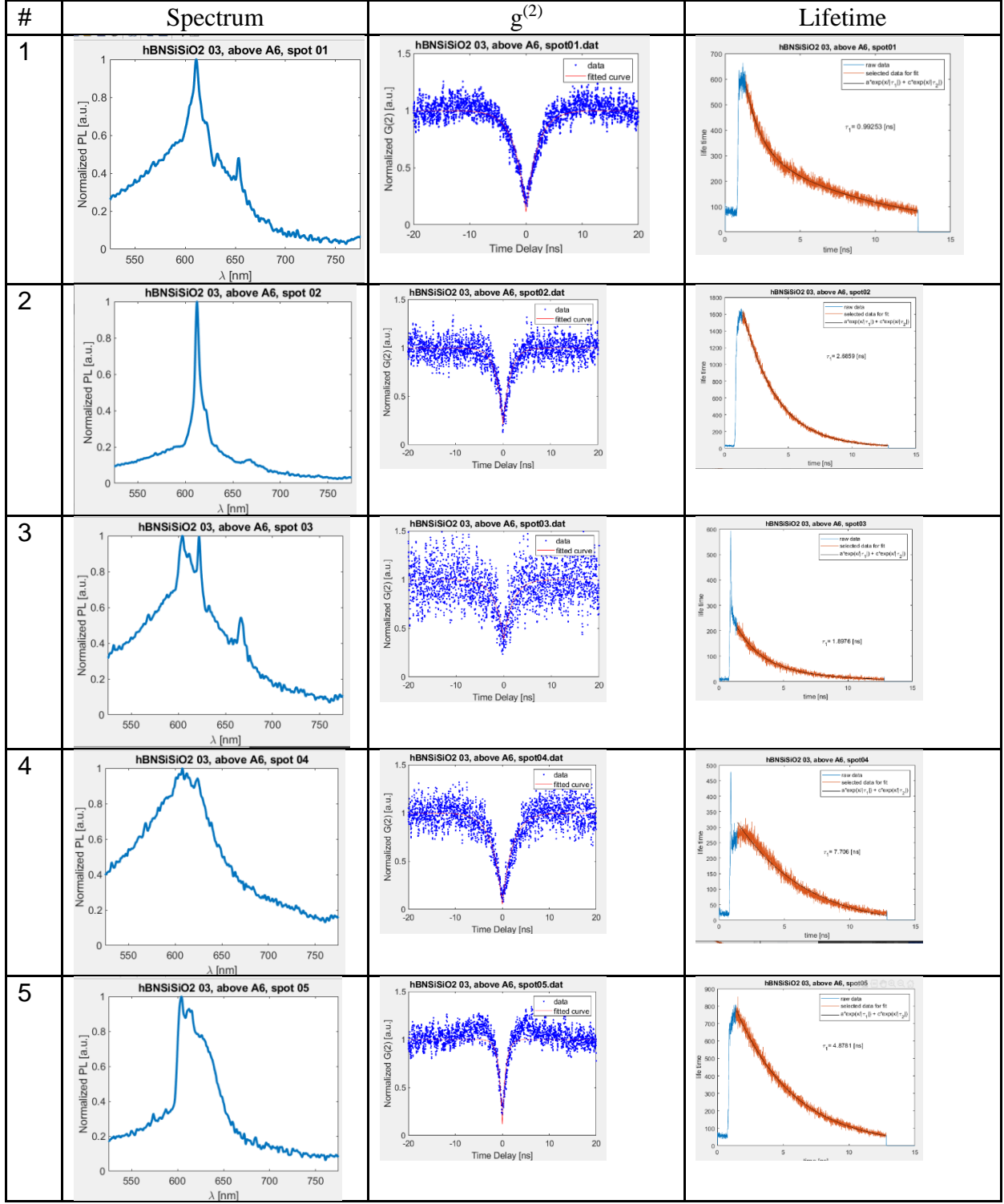

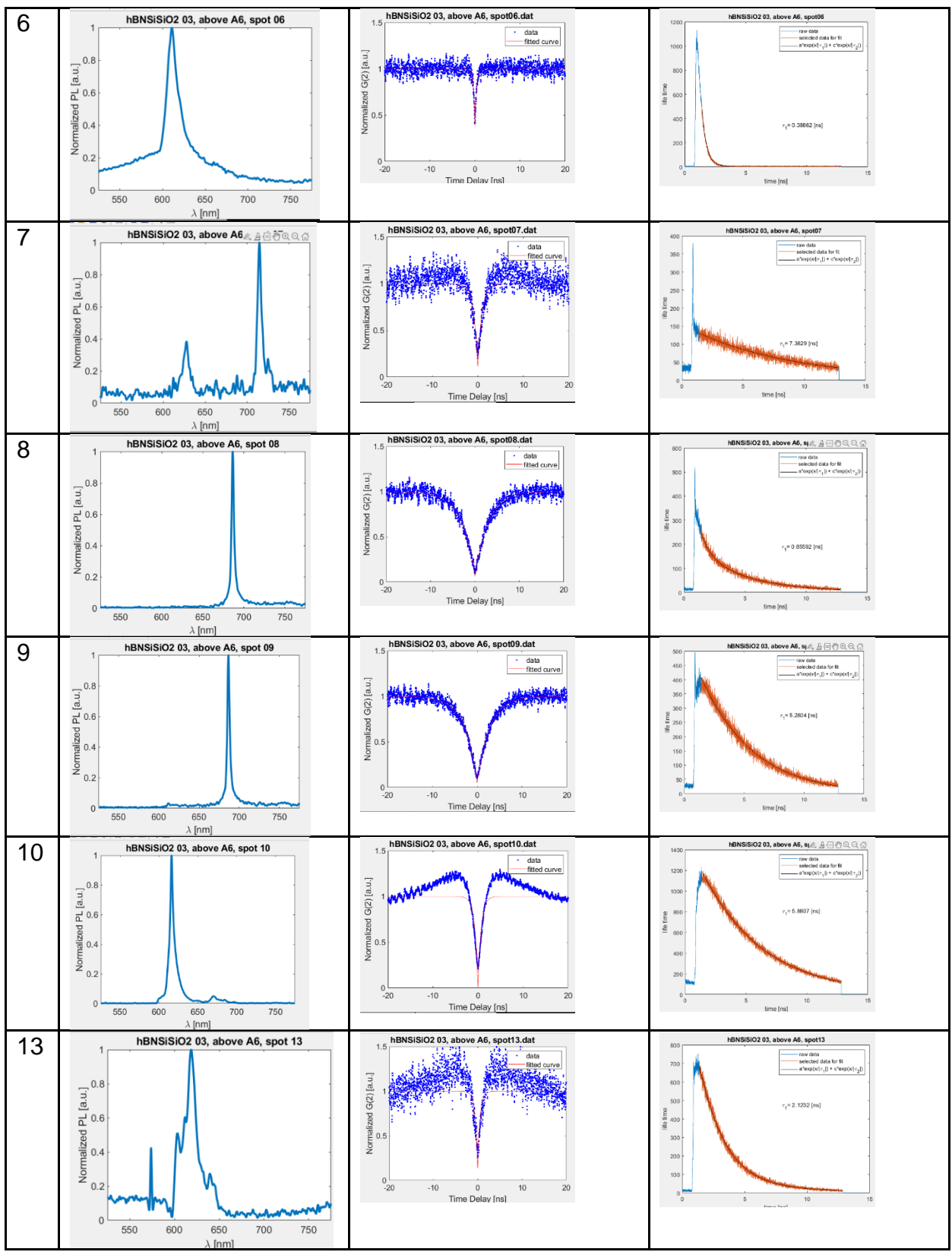

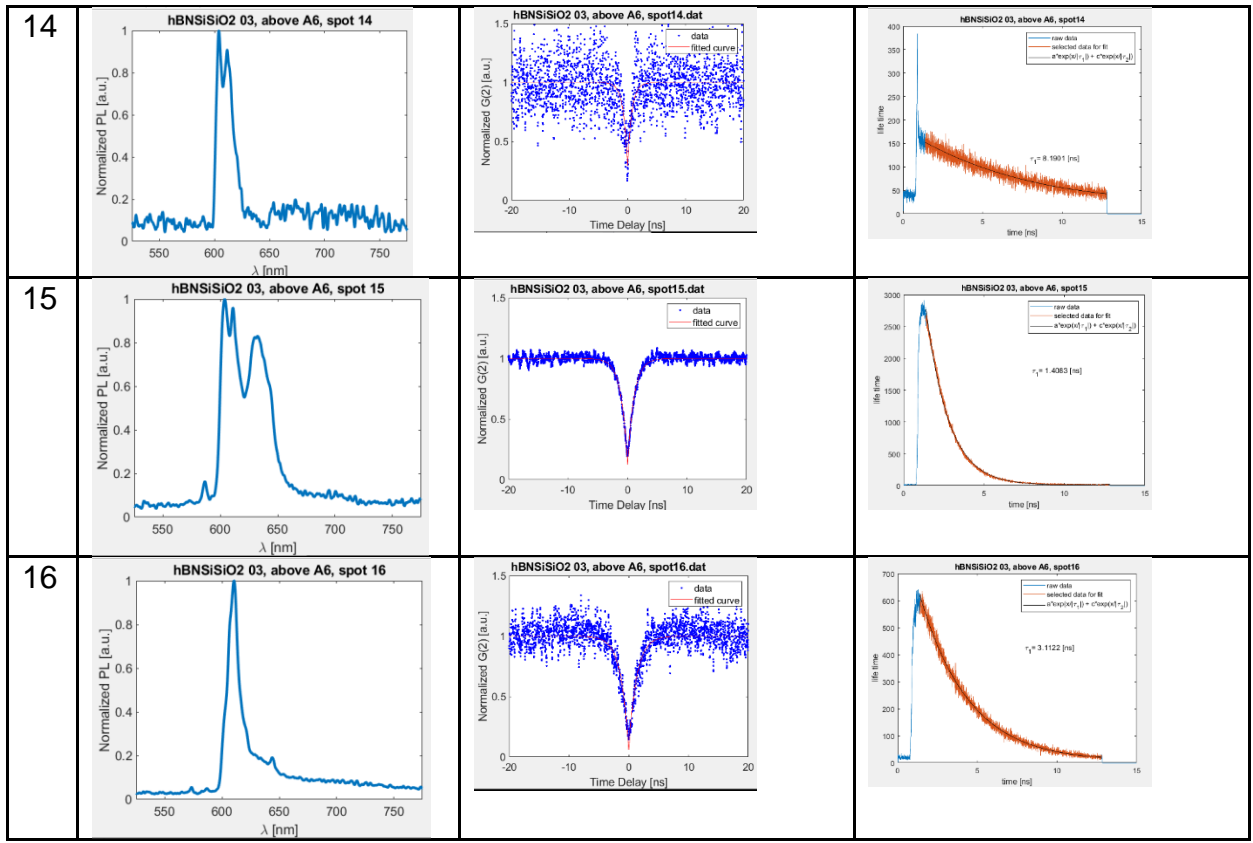

## V. Silver nanocube spreading

Silver nanocubes (SNCs) of 98 nm ±7 nm in size were commercially purchased from nanoCopisix and spin coated on the substrate. We optimized the spin coating process to increase the chance of getting the SNCs coupled to the targeted SPEs on hBN flake 1. Before spin coating, we sonicated the SNC solution for 5 minutes at room temperature to prevent clustering of the nanocubes and make a uniform solution. Then we placed 2 μL of SNC solution at the center of the $SiO_2$/Si sample on the spin coater. We set it to spin 300 rpm for 50 sec, followed by 3000 rpm for 10 sec. We verified the distribution of the SNCs on the hBN flake by optical microscopy: before (**Fig. S4a**) and after (**Fig. S4b**) spin coating. We also performed an AFM analysis of the SNCs (**Fig. 3a**) and confirmed the size of the cubes in the range of 95 to 105 nm (**Fig. 3b**).

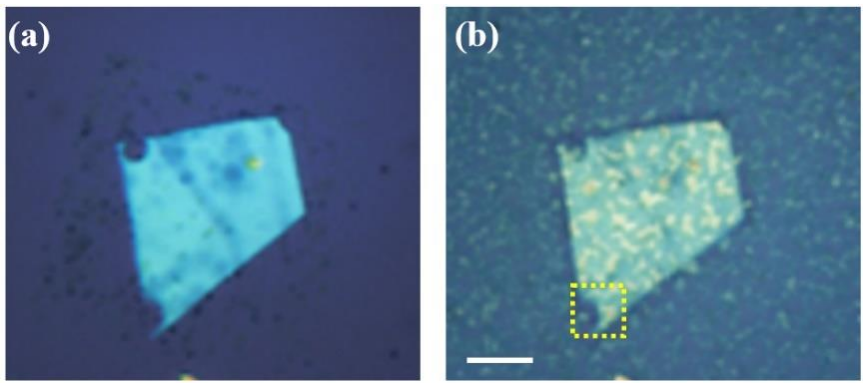

**Figure S4:** Optical image of the flake before (**a**) and after (**b**) spin coating of the SNCs on top of the hBN flake. The scale bar in (**a**) and (**b**) is 10 μm.

We characterized the optical properties of the SNCs by using the confocal setup in **Fig. 1b**. **Fig. S5a** shows the normalized autofluorescence spectrum from an individual ~ 98 nm SNC outside the flake. The high autofluorescence background is confirmed by scanning a region of the hBN flake with SNCs and SPEs (**Fig. S5b**) with a count rate at saturation of up to 5 x $10^6$ c/s. All SPE measurements performed after spreading the SNCs are subtracted from the hBN and SNC the autofluorescence background.

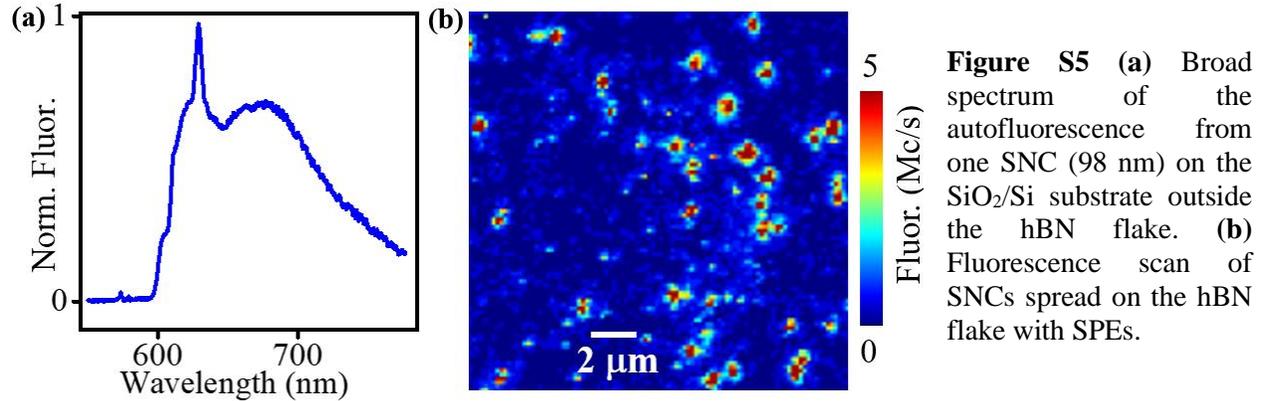

**Figure S5 (a)** Broad spectrum of the autofluorescence from one SNC (98 nm) on the SiO$_2$/Si substrate outside the hBN flake. **(b)** Fluorescence scan of SNCs spread on the hBN flake with SPEs.

## IV. Plasmon enhancement of quantum properties of emitter SPE 15

We measured the SPE quantum properties of emitter 16 before and after spin-coating of 98 nm ± 7 nm SNCs on top of the hBN flake in the insert of **Fig. 2a**. We see a plasmonic enhancement of the quantum properties of emitter SPE 15, i.e., a decrease of emission FWHM by 10% (**Fig. S6a**) and a narrowing of the $g^{(2)}$ response in **Fig. S6b** confirmed by the decrease of lifetime of 58% from 1.48 ns ± 0.075 ns to 0.87 ns ± 0.044 ns (**Fig. S6c**).

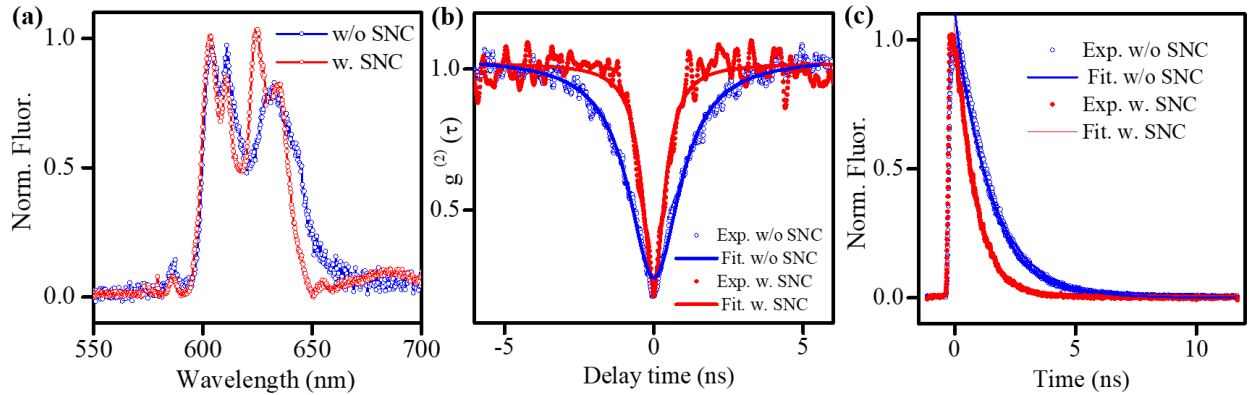

**Figure S6:** The quantum properties of emitter SPE 15 on the hBN flake are enhanced by coupling it to the SNC: **(a)** narrowing of the emission spectra, **(b)** narrowing of the autocorrelation response, and **(c)** shortening of the lifetime. In all these plots, open-circles (filled-circles) traces correspond to the measurements without (with) SNCs. Measurement settings are the same as the measurements presented in the main manuscript (**Fig. 3**).

**References**

(1) Aharonovich, I.; Englund, D.; Toth, M. Solid-State Single-Photon Emitters. *Nature Photon* **2016**, *10* (10), 631–641. https://doi.org/10.1038/nphoton.2016.186.

(2) Senellart, P.; Solomon, G.; White, A. High-Performance Semiconductor Quantum-Dot Single-Photon Sources. *Nature Nanotech* **2017**, *12* (11), 1026–1039. https://doi.org/10.1038/nnano.2017.218.

(3) Acosta, V.; Hemmer, P. Nitrogen-Vacancy Centers: Physics and Applications. *MRS Bull.* **2013**, *38* (2), 127–130. https://doi.org/10.1557/mrs.2013.18.

(4) Atatüre, M.; Englund, D.; Vamivakas, N.; Lee, S.-Y.; Wrachtrup, J. Material Platforms for Spin-Based Photonic Quantum Technologies. *Nat Rev Mater* **2018**, *3* (5), 38–51. https://doi.org/10.1038/s41578-018-0008-9.

(5) Laraoui, A.; Hodges, J. S.; Meriles, C. A. Nitrogen-Vacancy-Assisted Magnetometry of Paramagnetic Centers in an Individual Diamond Nanocrystal. *Nano Lett.* **2012**, *12* (7), 3477–3482. https://doi.org/10.1021/nl300964g.

(6) Trusheim, M. E.; Li, L.; Laraoui, A.; Chen, E. H.; Bakhru, H.; Schröder, T.; Gaathon, O.; Meriles, C. A.; Englund, D. Scalable Fabrication of High Purity Diamond Nanocrystals with Long-Spin-Coherence Nitrogen Vacancy Centers. *Nano Lett.* **2014**, *14* (1), 32–36. https://doi.org/10.1021/nl402799u.

(7) Laraoui, A.; Ambal, K. Opportunities for Nitrogen-Vacancy-Assisted Magnetometry to Study Magnetism in 2D van Der Waals Magnets. *Appl. Phys. Lett.* **2022**, *121* (6), 060502. https://doi.org/10.1063/5.0091931.

(8) Erickson, A.; Shah, S. Q. A.; Mahmood, A.; Fescenko, I.; Timalsina, R.; Binek, C.; Laraoui, A. Nanoscale Imaging of Antiferromagnetic Domains in Epitaxial Films of $Cr_2O_3$ via Scanning Diamond Magnetic Probe Microscopy. *RSC Advances* **2023**, *13* (1), 178–185. https://doi.org/10.1039/D2RA06440E.

(9) Diler, B.; Whiteley, S. J.; Anderson, C. P.; Wolfowicz, G.; Wesson, M. E.; Bielejec, E. S.; Joseph Heremans, F.; Awschalom, D. D. Coherent Control and High-Fidelity Readout of Chromium Ions in Commercial Silicon Carbide. *npj Quantum Inf* **2020**, *6* (1), 1–6. https://doi.org/10.1038/s41534-020-0247-7.

(10) Christle, D. J.; Falk, A. L.; Andrich, P.; Klimov, P. V.; Hassan, J. U.; Son, N. T.; Janzén, E.; Ohshima, T.; Awschalom, D. D. Isolated Electron Spins in Silicon Carbide with Millisecond Coherence Times. *Nature Mater* **2015**, *14* (2), 160–163. https://doi.org/10.1038/nmat4144.

(11) Wang, J.-F.; Li, Q.; Yan, F.-F.; Liu, H.; Guo, G.-P.; Zhang, W.-P.; Zhou, X.; Guo, L.-P.; Lin, Z.-H.; Cui, J.-M.; Xu, X.-Y.; Xu, J.-S.; Li, C.-F.; Guo, G.-C. On-Demand Generation of Single Silicon Vacancy Defects in Silicon Carbide. *ACS Photonics* **2019**, *6* (7), 1736–1743. https://doi.org/10.1021/acsphotonics.9b00451.

(12) Koehl, W. F.; Buckley, B. B.; Heremans, F. J.; Calusine, G.; Awschalom, D. D. Room Temperature Coherent Control of Defect Spin Qubits in Silicon Carbide. *Nature* **2011**, *479* (7371), 84–87. https://doi.org/10.1038/nature10562.

(13) Parto, K.; Azzam, S. I.; Banerjee, K.; Moody, G. Defect and Strain Engineering of Monolayer $WSe_2$ Enables Site-Controlled Single-Photon Emission up to 150 K. *Nat Commun* **2021**, *12* (1), 3585. https://doi.org/10.1038/s41467-021-23709-5.

(14) Klein, J.; Lorke, M.; Florian, M.; Sigger, F.; Sigl, L.; Rey, S.; Wierzbowski, J.; Cerne, J.; Müller, K.; Mitterreiter, E.; Zimmermann, P.; Taniguchi, T.; Watanabe, K.; Wurstbauer, U.; Kaniber, M.; Knap, M.; Schmidt, R.; Finley, J. J.; Holleitner, A. W. Site-Selectively Generated Photon Emitters in Monolayer $MoS_2$ via Local Helium Ion Irradiation. *Nat Commun* **2019**, *10* (1), 2755. https://doi.org/10.1038/s41467-019-10632-z.

(15) Eisaman, M. D.; Fan, J.; Migdall, A.; Polyakov, S. V. Invited Review Article: Single-Photon Sources and Detectors. *Review of Scientific Instruments* **2011**, *82* (7), 071101. https://doi.org/10.1063/1.3610677.

(16) Aharonovich, I.; Tetienne, J.-P.; Toth, M. Quantum Emitters in Hexagonal Boron Nitride. *Nano Lett.* **2022**, *22* (23), 9227–9235. https://doi.org/10.1021/acs.nanolett.2c03743.

(17) Tran, T. T.; Zachreson, C.; Berhane, A. M.; Bray, K.; Sandstrom, R. G.; Li, L. H.; Taniguchi, T.; Watanabe, K.; Aharonovich, I.; Toth, M. Quantum Emission from Defects in Single-Crystalline Hexagonal Boron Nitride. *Phys. Rev. Applied* **2016**, *5* (3), 034005. https://doi.org/10.1103/PhysRevApplied.5.034005.

(18) Tran, T. T.; Bray, K.; Ford, M. J.; Toth, M.; Aharonovich, I. Quantum Emission from Hexagonal Boron Nitride Monolayers. *Nature Nanotech* **2016**, *11* (1), 37–41. https://doi.org/10.1038/nnano.2015.242.

(19) Tran, T. T.; Elbadawi, C.; Totonjian, D.; Lobo, C. J.; Grosso, G.; Moon, H.; Englund, D. R.; Ford, M. J.; Aharonovich, I.; Toth, M. Robust Multicolor Single Photon Emission from Point Defects in Hexagonal Boron Nitride. *ACS Nano* **2016**, *10* (8), 7331–7338. https://doi.org/10.1021/acsnano.6b03602.

(20) Xu, Z.-Q.; Elbadawi, C.; Tran, T. T.; Kianinia, M.; Li, X.; Liu, D.; Hoffman, T. B.; Nguyen, M.; Kim, S.; Edgar, J. H.; Wu, X.; Song, L.; Ali, S.; Ford, M.; Toth, M.; Aharonovich, I. Single Photon Emission from Plasma Treated 2D Hexagonal Boron Nitride. *Nanoscale* **2018**, *10* (17), 7957–7965. https://doi.org/10.1039/C7NR08222C.

(21) Li, C.; Xu, Z.-Q.; Mendelson, N.; Kianinia, M.; Toth, M.; Aharonovich, I. Purification of Single-Photon Emission from HBN Using Post-Processing Treatments. *Nanophotonics* **2019**, *8* (11), 2049–2055. https://doi.org/10.1515/nanoph-2019-0099.

(22) Gao, T.; von Helversen, M.; Antón-Solanas, C.; Schneider, C.; Heindel, T. Atomically-Thin Single-Photon Sources for Quantum Communication. *npj 2D Mater Appl* **2023**, *7* (1), 1–9. https://doi.org/10.1038/s41699-023-00366-4.

(23) Zhang, C.; Shi, Z.; Wu, T.; Xie, X. Microstructure Engineering of Hexagonal Boron Nitride for Single-Photon Emitter Applications. *Advanced Optical Materials* **2022**, *10* (17), 2200207. https://doi.org/10.1002/adom.202200207.

(24) Stern, H. L.; Gu, Q.; Jarman, J.; Eizagirre Barker, S.; Mendelson, N.; Chugh, D.; Schott, S.; Tan, H. H.; Sirringhaus, H.; Aharonovich, I.; Atatüre, M. Room-Temperature Optically Detected Magnetic Resonance of Single Defects in Hexagonal Boron Nitride. *Nat Commun* **2022**, *13* (1), 618. https://doi.org/10.1038/s41467-022-28169-z.

(25) Caldwell, J. D.; Aharonovich, I.; Cassabois, G.; Edgar, J. H.; Gil, B.; Basov, D. N. Photonics with Hexagonal Boron Nitride. *Nat Rev Mater* **2019**, *4* (8), 552–567. https://doi.org/10.1038/s41578-019-0124-1.

(26) Toth, M.; Aharonovich, I. Single Photon Sources in Atomically Thin Materials. *Annu. Rev. Phys. Chem.* **2019**, *70* (1), 123–142. https://doi.org/10.1146/annurev-physchem-042018-052628.

(27) Degen, C. L.; Reinhard, F.; Cappellaro, P. Quantum Sensing. *Rev. Mod. Phys.* **2017**, *89* (3), 035002. https://doi.org/10.1103/RevModPhys.89.035002.

(28) Grosso, G.; Moon, H.; Lienhard, B.; Ali, S.; Efetov, D. K.; Furchi, M. M.; Jarillo-Herrero, P.; Ford, M. J.; Aharonovich, I.; Englund, D. Tunable and High-Purity Room Temperature Single-Photon Emission from Atomic Defects in Hexagonal Boron Nitride. *Nat Commun* **2017**, *8* (1), 705. https://doi.org/10.1038/s41467-017-00810-2.

(29) Nikolay, N.; Mendelson, N.; Özelci, E.; Sontheimer, B.; Böhm, F.; Kewes, G.; Toth, M.; Aharonovich, I.; Benson, O. Direct Measurement of Quantum Efficiency of Single-Photon Emitters in Hexagonal Boron Nitride. *Optica* **2019**, *6* (8), 1084. https://doi.org/10.1364/OPTICA.6.001084.

(30) Mendelson, N.; Xu, Z.-Q.; Tran, T. T.; Kianinia, M.; Scott, J.; Bradac, C.; Aharonovich, I.; Toth, M. Engineering and Tuning of Quantum Emitters in Few-Layer Hexagonal Boron Nitride. *ACS Nano* **2019**, *13* (3), 3132–3140. https://doi.org/10.1021/acsnano.8b08511.

(31) Liu, H.; You, C. Y.; Li, J.; Galligan, P. R.; You, J.; Liu, Z.; Cai, Y.; Luo, Z. Synthesis of Hexagonal Boron Nitrides by Chemical Vapor Deposition and Their Use as Single Photon Emitters. *Nano Materials Science* **2021**, *3* (3), 291–312. https://doi.org/10.1016/j.nanoms.2021.03.002.

(32) Boll, M. K.; Radko, I. P.; Huck, A.; Andersen, U. L. Photophysics of Quantum Emitters in Hexagonal Boron-Nitride Nano-Flakes. *Opt. Express* **2020**, *28* (5), 7475. https://doi.org/10.1364/OE.386629.

(33) Chen, Y.; Li, C.; White, S.; Nonahal, M.; Xu, Z.-Q.; Watanabe, K.; Taniguchi, T.; Toth, M.; Tran, T. T.; Aharonovich, I. Generation of High-Density Quantum Emitters in High-Quality, Exfoliated Hexagonal Boron Nitride. *ACS Appl. Mater. Interfaces* **2021**, *13* (39), 47283–47292. https://doi.org/10.1021/acsami.1c14863.

(34) Xu, X.; Martin, Z. O.; Sychev, D.; Lagutchev, A. S.; Chen, Y. P.; Taniguchi, T.; Watanabe, K.; Shalaev, V. M.; Boltasseva, A. Creating Quantum Emitters in Hexagonal Boron Nitride Deterministically on Chip-Compatible Substrates. *Nano Lett.* **2021**, *21* (19), 8182–8189. https://doi.org/10.1021/acs.nanolett.1c02640.

(35) Mendelson, N.; Chugh, D.; Reimers, J. R.; Cheng, T. S.; Gottscholl, A.; Long, H.; Mellor, C. J.; Zettl, A.; Dyakonov, V.; Beton, P. H.; Novikov, S. V.; Jagadish, C.; Tan, H. H.; Ford, M. J.; Toth, M.; Bradac, C.; Aharonovich, I. Identifying Carbon as the Source of Visible Single-Photon Emission from Hexagonal Boron Nitride. *Nat. Mater.* **2021**, *20* (3), 321–328. https://doi.org/10.1038/s41563-020-00850-y.

(36) Englund, D.; Shields, B.; Rivoire, K.; Hatami, F.; Vučković, J.; Park, H.; Lukin, M. D. Deterministic Coupling of a Single Nitrogen Vacancy Center to a Photonic Crystal Cavity. *Nano Lett.* **2010**, *10* (10), 3922–3926. https://doi.org/10.1021/nl101662v.

(37) Faraon, A.; Santori, C.; Huang, Z.; Acosta, V. M.; Beausoleil, R. G. Coupling of Nitrogen-Vacancy Centers to Photonic Crystal Cavities in Monocrystalline Diamond. *Phys. Rev. Lett.* **2012**, *109* (3), 033604. https://doi.org/10.1103/PhysRevLett.109.033604.

(38) Li, L.; Chen, E. H.; Zheng, J.; Mouradian, S. L.; Dolde, F.; Schröder, T.; Karaveli, S.; Markham, M. L.; Twitchen, D. J.; Englund, D. Efficient Photon Collection from a Nitrogen Vacancy Center in a Circular Bullseye Grating. *Nano Lett.* **2015**, *15* (3), 1493–1497. https://doi.org/10.1021/nl503451j.

(39) Yang, G.; Shen, Q.; Niu, Y.; Wei, H.; Bai, B.; Mikkelsen, M. H.; Sun, H. Unidirectional, Ultrafast, and Bright Spontaneous Emission Source Enabled By a Hybrid Plasmonic Nanoantenna. *Laser & Photonics Reviews* **2020**, *14* (3), 1900213. https://doi.org/10.1002/lpor.201900213.

(40) Bogdanov, S. I.; Shalaginov, M. Y.; Lagutchev, A. S.; Chiang, C.-C.; Shah, D.; Baburin, A. S.; Ryzhikov, I. A.; Rodionov, I. A.; Kildishev, A. V.; Boltasseva, A.; Shalaev, V. M. Ultrabright Room-Temperature Sub-Nanosecond Emission from Single Nitrogen-Vacancy Centers Coupled to Nanopatch Antennas. *Nano Lett.* **2018**, *18* (8), 4837–4844. https://doi.org/10.1021/acs.nanolett.8b01415.

(41) Preuß, J. A.; Rudi, E.; Kern, J.; Schmidt, R.; Bratschitsch, R.; Michaelis de Vasconcellos, S. Assembly of Large HBN Nanocrystal Arrays for Quantum Light Emission. *2D Mater.* **2021**, *8* (3), 035005. https://doi.org/10.1088/2053-1583/abeca2.

(42) Kim, S.; Fröch, J. E.; Christian, J.; Straw, M.; Bishop, J.; Totonjian, D.; Watanabe, K.; Taniguchi, T.; Toth, M.; Aharonovich, I. Photonic Crystal Cavities from Hexagonal Boron Nitride. *Nat Commun* **2018**, *9* (1), 2623. https://doi.org/10.1038/s41467-018-05117-4.

(43) Mendelson, N.; Ritika, R.; Kianinia, M.; Scott, J.; Kim, S.; Fröch, J. E.; Gazzana, C.; Westerhausen, M.; Xiao, L.; Mohajerani, S. S.; Strauf, S.; Toth, M.; Aharonovich, I.; Xu, Z. Coupling Spin Defects in a Layered Material to Nanoscale Plasmonic Cavities. *Advanced Materials* **2022**, *34* (1), 2106046. https://doi.org/10.1002/adma.202106046.

(44) Xu, X.; Solanki, A. B.; Sychev, D.; Gao, X.; Peana, S.; Baburin, A. S.; Pagadala, K.; Martin, Z. O.; Chowdhury, S. N.; Chen, Y. P.; Taniguchi, T.; Watanabe, K.; Rodionov, I. A.; Kildishev, A. V.; Li, T.; Upadhyaya, P.; Boltasseva, A.; Shalaev, V. M. Greatly Enhanced Emission from Spin Defects in Hexagonal Boron Nitride Enabled by a Low-Loss Plasmonic Nanocavity. *Nano Lett.* **2022**, acs.nanolett.2c03100. https://doi.org/10.1021/acs.nanolett.2c03100.

(45) Hoang, T. B.; Akselrod, G. M.; Argyropoulos, C.; Huang, J.; Smith, D. R.; Mikkelsen, M. H. Ultrafast Spontaneous Emission Source Using Plasmonic Nanoantennas. *Nat Commun* **2015**, *6* (1), 7788. https://doi.org/10.1038/ncomms8788.

(46) Akselrod, G. M.; Argyropoulos, C.; Hoang, T. B.; Ciracì, C.; Fang, C.; Huang, J.; Smith, D. R.; Mikkelsen, M. H. Probing the Mechanisms of Large Purcell Enhancement in Plasmonic Nanoantennas. *Nature Photon* **2014**, *8* (11), 835–840. https://doi.org/10.1038/nphoton.2014.228.

(47) Iff, O.; Buchinger, Q.; Moczała-Dusanowska, M.; Kamp, M.; Betzold, S.; Davanco, M.; Srinivasan, K.; Tongay, S.; Antón-Solanas, C.; Höfling, S.; Schneider, C. Purcell-Enhanced Single Photon Source Based on a Deterministically Placed WSe$_2$ Monolayer Quantum Dot in a Circular Bragg Grating Cavity. *Nano Lett.* **2021**, *21* (11), 4715–4720. https://doi.org/10.1021/acs.nanolett.1c00978.

(48) Healey, A. J.; Scholten, S. C.; Yang, T.; Scott, J. A.; Abrahams, G. J.; Robertson, I. O.; Hou, X. F.; Guo, Y. F.; Rahman, S.; Lu, Y.; Kianinia, M.; Aharonovich, I.; Tetienne, J.-P. Quantum Microscopy with van Der Waals Heterostructures. *Nat. Phys.* **2023**, *19* (1), 87–91. https://doi.org/10.1038/s41567-022-01815-5.

(49) Huang, M.; Zhou, J.; Chen, D.; Lu, H.; McLaughlin, N. J.; Li, S.; Alghamdi, M.; Djugba, D.; Shi, J.; Wang, H.; Du, C. R. Wide Field Imaging of van Der Waals Ferromagnet Fe3GeTe2 by Spin Defects in Hexagonal Boron Nitride. *Nat Commun* **2022**, *13* (1), 5369. https://doi.org/10.1038/s41467-022-33016-2.

(50) Bogdanov, S. I.; Boltasseva, A.; Shalaev, V. M. Overcoming Quantum Decoherence with Plasmonics. *Science* **2019**, *364* (6440), 532–533. https://doi.org/10.1126/science.aax3766.

(51) Ares, P.; Santos, H.; Lazić, S.; Gibaja, C.; Torres, I.; Pinilla, S.; Gómez-Herrero, J.; Meulen, H. P.; García-González, P.; Zamora, F. Direct Visualization and Effects of Atomic-Scale Defects on the Optoelectronic Properties of Hexagonal Boron Nitride. *Adv. Electron. Mater.* **2021**, *7* (7), 2001177. https://doi.org/10.1002/aelm.202001177.

(52) Weston, L.; Wickramaratne, D.; Mackoit, M.; Alkauskas, A.; Van de Walle, C. G. Native Point Defects and Impurities in Hexagonal Boron Nitride. *Phys. Rev. B* **2018**, *97* (21), 214104. https://doi.org/10.1103/PhysRevB.97.214104.

(53) Anzai, Y.; Yamamoto, M.; Genchi, S.; Watanabe, K.; Taniguchi, T.; Ichikawa, S.; Fujiwara, Y.; Tanaka, H. Broad Range Thickness Identification of Hexagonal Boron Nitride by Colors. *Appl. Phys. Express* **2019**, *12* (5), 055007. https://doi.org/10.7567/1882-0786/ab0e45.

(54) Kitson, S. C.; Jonsson, P.; Rarity, J. G.; Tapster, P. R. Intensity Fluctuation Spectroscopy of Small Numbers of Dye Molecules in a Microcavity. *Phys. Rev. A* **1998**, *58* (1), 620–627. https://doi.org/10.1103/PhysRevA.58.620.

(55) Wu, E.; Jacques, V.; Zeng, H.; Grangier, P.; Treussart, F.; Roch, J.-F. Narrow-Band Single-Photon Emission in the near Infrared for Quantum Key Distribution. *Opt. Express* **2006**, *14* (3), 1296. https://doi.org/10.1364/OE.14.001296.

(56) Akselrod, G. M.; Ming, T.; Argyropoulos, C.; Hoang, T. B.; Lin, Y.; Ling, X.; Smith, D. R.; Kong, J.; Mikkelsen, M. H. Leveraging Nanocavity Harmonics for Control of Optical Processes in 2D Semiconductors. *Nano Lett.* **2015**, *15* (5), 3578–3584. https://doi.org/10.1021/acs.nanolett.5b01062.

(57) Ciracì, C.; Rose, A.; Argyropoulos, C.; Smith, D. R. Numerical Studies of the Modification of Photodynamic Processes by Film-Coupled Plasmonic Nanoparticles. *J. Opt. Soc. Am. B* **2014**, *31* (11), 2601. https://doi.org/10.1364/JOSAB.31.002601.

(58) Nguyen, M.; Kim, S.; Tran, T. T.; Xu, Z.-Q.; Kianinia, M.; Toth, M.; Aharonovich, I. Nanoassembly of Quantum Emitters in Hexagonal Boron Nitride and Gold Nanospheres. *Nanoscale* **2018**, *10* (5), 2267–2274. https://doi.org/10.1039/C7NR08249E.

(59) Rah, Y.; Jin, Y.; Kim, S.; Yu, K. Optical Analysis of the Refractive Index and Birefringence of Hexagonal Boron Nitride from the Visible to Near-Infrared. *Opt. Lett.* **2019**, *44* (15), 3797. https://doi.org/10.1364/OL.44.003797.

(60) Johnson, P. B.; Christy, R. W. Optical Constants of the Noble Metals. *Phys. Rev. B* **1972**, *6* (12), 4370–4379. https://doi.org/10.1103/PhysRevB.6.4370.

(61) Rodríguez-de Marcos, L. V.; Larruquert, J. I.; Méndez, J. A.; Aznárez, J. A. Self-Consistent Optical Constants of SiO_2 and Ta_2O_5 Films. *Opt. Mater. Express* **2016**, *6* (11), 3622. https://doi.org/10.1364/OME.6.003622.

(62) Schinke, C.; Christian Peest, P.; Schmidt, J.; Brendel, R.; Bothe, K.; Vogt, M. R.; Kröger, I.; Winter, S.; Schirmacher, A.; Lim, S.; Nguyen, H. T.; MacDonald, D. Uncertainty Analysis for the Coefficient of Band-to-Band Absorption of Crystalline Silicon. *AIP Advances* **2015**, *5* (6), 067168. https://doi.org/10.1063/1.4923379.

(63) Ren, S.; Tan, Q.; Zhang, J. Review on the Quantum Emitters in Two-Dimensional Materials. *J. Semicond.* **2019**, *40* (7), 071903. https://doi.org/10.1088/1674-4926/40/7/071903.